\documentclass[a4paper,12pt]{article}
\usepackage[utf8]{inputenc}
\usepackage{amsmath,amsthm,,amsfonts,amssymb}
\usepackage{booktabs,tabulary}
\usepackage{array}
\usepackage{bm}
\usepackage{multirow}
\usepackage{pdfpages}
\usepackage{enumitem}
\usepackage{siunitx}
\usepackage{caption}

\usepackage[margin=2.9cm]{geometry}
\usepackage{tikz}
\usepackage{tkz-graph}
\usetikzlibrary{calc}
\usetikzlibrary{shapes.geometric}%
\RequirePackage[colorlinks,citecolor=blue,urlcolor=blue]{hyperref}

\DeclareMathOperator*{\argmin}{\arg\!\min}		
\newcommand\pb{\bm{p}}
\newcommand{\pr}{\mathbb{P}}
\newcommand{\E}{\mathbb{E}}

\newcommand\zb{\bm{z}}
\newcommand\nb{\bm{n}}
\newcommand\alphab{\bm{\alpha}}

\makeatletter
\newcommand*{\rom}[1]{\expandafter\@slowromancap\romannumeral #1@}
\makeatother
\newcommand\betab{\bm{\beta}}

\DeclareMathOperator{\mtd}{MTD}
\usepackage{natbib}	
\usepackage{algorithm}
\usepackage{algpseudocode}
\usepackage{xpatch}
\makeatletter
\xpatchcmd{\algorithmic}{\itemsep\z@}{\itemsep=.5ex plus2pt}{}{}
\makeatother

\RequirePackage{amsthm}
\theoremstyle{plain}

\newtheorem*{thm*}{Theorem}
\newtheorem*{prop*}{Proposition}
\newtheorem*{cor*}{Corollary}
\newtheorem*{lem*}{Lemma}
\newtheorem*{assum*}{Assumption}

\theoremstyle{definition}
\newtheorem{defn}{Definition}

\newtheorem*{defn*}{Definition}
\newtheorem*{rem*}{Remark}
\newtheorem*{exa*}{Example}

\usepackage{tabularx}

\author{Zahra S. Razaee, Galen Wien-Cook, Mourad Tighiouart}
\begin{document}
\title{A Nonparametric Bayesian Design for Drug Combination Cancer Trials}

\maketitle
\begin{abstract}
	We propose an adaptive design for early phase drug combination cancer trials with the goal of
	estimating the maximum tolerated dose (MTD). A nonparametric Bayesian model, using beta
	priors truncated to the set of partially ordered dose combinations, is used to describe the
	probability of dose limiting toxicity (DLT). Dose allocation between successive cohorts of patients is estimated using a modified Continual Reassessment scheme. The updated probabilities of DLT are calculated with a Gibbs sampler that employs
	a weighting mechanism to calibrate the influence of data
	versus the prior. 
	 At the end of the trial, we recommend one or more dose combinations as the MTD
	 based on our proposed algorithm. 
	 The design operating characteristics indicate that our method is comparable with existing methods. As an illustration, we apply our method to a phase \rom{1} clinical trial of CB-839 and Gemcitabine. 
\end{abstract}

Cancer phase I trials, Drug combination, Maximum tolerated dose, Nonparametric Bayesian design, Partial ordering. 
\section{Introduction}
The primary objective in conventional phase \rom{1} clinical trials is to determine the maximum tolerated dose (MTD), defined as the dose with the probability of toxicity closest to a  prespecified target. For safety and ethical concerns, most phase \rom{1} trials are conducted adaptively, using the dose limiting toxicity (DLT) status of  previously enrolled patients to determine the dose level for the next cohort of patients. The majority of such trials are designed for single agent, e.g., the conventional 3 + 3 design \citep{storer1989design}, the continual reassessment method(CRM) and its variants \citep{o1990continual}, \citep{goodman1995some}, \citep{korn1991selecting}, \citep{moller1995extension}, \citep{o1996continual}, 
\citep{leung2002extension}, \citep{o2003continual}, \citep{iasonos2011continual}, \citep{daimon2011posterior}, \citep{liu2013Bayesian}, the efficient dose escalation with overdose control (EWOC) method and its variants \citep{babb1998cancer,tighiouart2014dose,tighiouart2010dose,wheeler2017toxicity}, the modified toxicity probability interval method \citep{ji2013modified}, the Bayesian optimal design \citep{yuan2016Bayesian}, the nonparametric overdose control method \citep{lin2017nonparametric}, the semiparametric dose finding methods \citep{clertant2017semiparametric} and the Bayesian adaptive design  using a flexible range of doses \citep{tighiouart2018bayesian}. 

Recent advances in drug discovery have intensified interest 
in using dual agents in phase \rom{1} clinical trials. This interest is fueled by the fact that drug combinations may induce a synergistic treatment effect by targeting multiple pathways simultaneously and inhibiting resistance mechanisms. 
A fundamental assumption for cytotoxic/biologic agents is monotonicity between toxicity and doses. 
For single agent, this assumption induces a complete ordering of the doses. However, in the case of drug combination treatment where the two agents are allowed to vary, it induces a partial ordering constraint on the probabilities of toxicities. The monotonicity assumption coupled with small sample size in phase \rom{1} clinical trials and higher dimension of the dose space, make the design of combination trials challenging.

Various model-based designs for drug combinations, have been studied in the last decade. \cite{thall2003dose} proposed a method using a six-parameter model to define the probability of toxicity as a function of the two doses with the requirement that  each of the two agents had been studied previously as a single agent. \cite{wang2005two} used a two-stage design with regression model. \cite{yin2009Bayesian} and \cite{yin2009latent} developed a design that models the probability of toxicity with a copula type model.
\cite{wages2011continual}  considered  estimation of toxicity probabilities within a small number of simple orders. 
\cite{tighiouart2017bayesian} and \cite{tighiouart2018bayesian} used a reparametrized logistic model to describe the relationship between the doses of the two agents and the probability of dose limiting toxicity and extended the work of \cite{tighiouart2014dose} by allowing the MTD curve to lie anywhere in the Cartesian plane of the dose levels of the two drugs and treating cohorts of two patients simultaneously with different dose combinations. The method was further extended to account for a baseline covariate by \cite{diniz2018bayesian} and settings where an unknown fraction of DLTs is attributable to one or more agents by \cite{jimenez2019cancer}.


In contrast to the parametric models, the nonparametric models do not impose any functional form on the dose–toxicity relationship. Parametric models suffer from potential model misspecification, which may lead to unsafe dose escalation.  On the other hand, nonparametric models can capture more subtle aspect of the data, hence they are more flexible. Several nonparametric  models for  dual agent clinical trials have been studied in the past. 
\cite{lin2016bootstrap} estimated the toxicity order of two drugs by two-dimensional isotonic regression and reduced the two-dimensional drug combination searching space into a one dimension and used a parametric CRM model based on the updated toxicity order.
 \cite{mander2015product} considered a product of independent beta probabilities escalation strategy allowing the prior distributions for each dose combination to be unconstrained and imposing the monotonicity assumption when escalating by choosing only monotonic contours. 

 In this paper, we propose a nonparametric 
 Bayesian approach to dose finding for combinations of drugs (NBCD) by modeling the joint prior probabilities of DLTs on the space of all dose combinations with independent beta distributions truncated to the set of combinations that satisfy the partial order. Unlike the PIPE algorithm proposed by \cite{mander2015product}, our approach guarantees that the joint posterior distribution of the probabilities of DLT estimated with Gibbs sampler satisfies the partial order constraint. A weighted mechanism is introduced when allocating doses to successive cohorts of patients in order to calibrate the influence of data and that of the prior.

 The rest of this paper is organized as follows. In Section~\ref{sec:model}, we propose a nonparametric Bayesian model with beta priors truncated to the set of partially ordered dose combinations. In Section~\ref{sec:trialdesign}, we present our trial design and an algorithm for  dose recommendation for phase \rom{2}.  We study the performance of our model with extensive simulations studies in Section~\ref{sec:simulations} and a trial of CB-839 and Gemcitabine in Section~\ref{sec:realtrial}.

\paragraph{Notation.}
We use $n_{ij}$ and $z_{ij}$ to denote the number of patient assigned to the dose combination $d_{ij}$ and the number of DLTs observed at $(i,j), respectively$. We use $\alphab = (\alpha_{ij}) \in \mathbb{R} ^K$ and $\betab = (\beta_{ij}) \in \mathbb{R} ^K$ that collect all $\alpha_{ij}$ and $\beta_{ij}$ parameters, respectively. For any positive integer $I$, we write $[I] = \{1,\dots,I\}$. We denote by $\tilde{p}_{ij}$ and $\hat{p}_{ij}$, the
prior and posterior median of the probability of DLT at dose level $d_{ij}$, respectively. 


\section{Problem formulation}
Consider two drugs $A$ and $B$ with ordered dose levels $\mathcal{A} = \{d_i^A , i = 1,\ldots,I \}$ and $\mathcal{B} = \{d_j^B , j= 1,\ldots,J \}$, respectively.  Let $\mathcal{D} = \mathcal{A} \times \mathcal{B}$ be 
 the set of dose combinations available in the trial.
We denote a typical element of $\mathcal D$ as $d_{ij} = (d_i^A, d_j^B)$. To each dose combination $d_{ij}$, there is a true probability of toxicity, or DLT, which we denote as $p^*_{ij}$. That is,
\begin{align}
	p^*_{ij} = \pr\big( \text{Dose combination $d_{ij}$ causes a DLT} \big).
\end{align}
Later in our model, we use $p_{ij}$ when modeling these probabilities as random. 

Given a target probability of DLT $\theta \in (0,1)$, we are interested in dose combinations whose $p^*_{ij}$ are close to $\theta$. In order to make this notion precise, we define $\delta$-approximate MTD as

\begin{align}
\text{MTD}(\delta) := \{ d_{ij} \mid (i,j) \in \mathcal{M}_\delta \}
\end{align}

where,
\begin{align}
\mathcal{M}_\delta = \big\{(i,j) \in [I] \times [J]:\;  |p^*_{ij} - \theta| \leq \delta \big\}
\end{align}

Our goal is to recover any dose combination in $\mtd(\delta)$ for prescribed values of $\delta$ and $\theta$. In the sequel, for simplicity, we often say dose combination $(i,j)$ instead of $d_{ij}$.

		
\subsection{The model}
When modeling drug combination trials, we consider the probabilities of DLT as random variables $p_{ij}$ for which we will specify a prior. Let $\pb = (p_{11},p_{12},p_{13},\dots,p_{IJ})$ be the random vector obtained by collecting all the $p_{ij}$s. During the trial, the patients are assigned to various dose combinations and their toxicity response is recorded. Assume that at a given stage in the trial, we have assigned a total of $n_{ij}$ patients to dose combination $(i,j)$. The number of these patients who experienced DLT, denoted as $z_{ij}$ is distributed as 
\begin{align*}
z_{ij} \mid \pb \sim \text{Bin}(n_{ij},p_{ij}).
\end{align*}
Letting $\mathcal N = \{(i,j) \mid n_{ij} > 0\}$, the likelihood of the model is
\begin{align}\label{eq:likelihood}
L(\zb \mid \pb) = \prod_{(i,j) \,\in\, \mathcal N} \binom{n_{ij}}{z_{ij}} p_{ij}^{z_{ij}}(1-p_{ij})^{n_{ij} - z_{ij}}.
\end{align}
where $\zb = (z_{ij})$ collects all the $z_{ij}$s.
In Section~\ref{subsec:npprior}, we specify a prior for $\pb$ which allows to obtain the posterior estimate of $\pb$ given $\zb = (z_{ij})$. Given these estimates, we update our estimate of $\mtd(\delta)$, assign more patients, and so on. Before specifying the prior on $\pb$, we need to better understand the constraints on $\pb$.

\begin{figure}[t] 
	\centering
	\includegraphics[width=.45\textwidth]{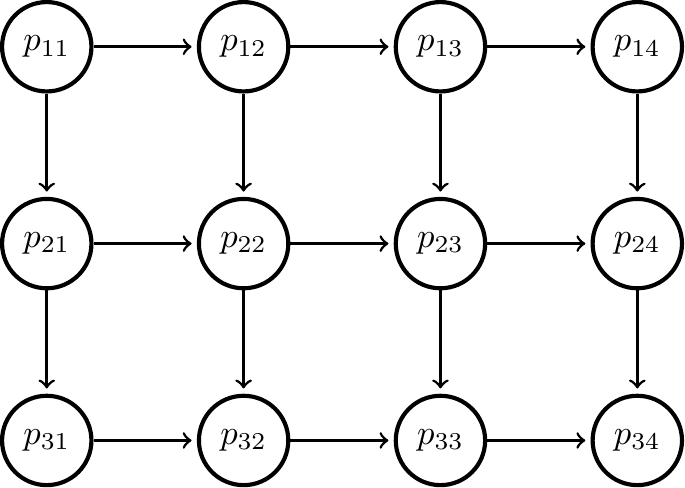}
	\caption{Hasse diagram for a $3\times 4$ lattice}
	\label{fig:hasse:diagram}
\end{figure}

\subsection{Lattice constraints}
We assume that the dose combination are ordered so that $p_{ij} \le p_{i'j'}$ if $ i \le i'$ and $j \le j'$. These constraints define a partial order on the collection $\{p_{ij},i=1,\dots,I, j=1,\dots,J\}$ which is illustrated using a Hasse diagram in Figure~\ref{fig:hasse:diagram}.

Recall that $\pb = (p_{11},p_{12},p_{13},\dots,p_{IJ})$ collect all the probabilities of DLT. The partial ordering constraints on $\pb$ can be encoded as the intersection of the following sets:
\begin{alignat}{3}
\Omega_1 & = \Omega_{11} = \bigl\{\,\pb \mid\; 0 &&< p_{11}&& < \min(p_{12},p_{21})\,\bigr\}, \nonumber\\
		 & \;\; \vdots\nonumber\\
\Omega_k& = \Omega_{ij} =  \bigl\{\,\pb \mid\; \max(p_{i-1,j},p_{i,j-1}) &&< p_{ij}&& < \min(p_{i,j+1},p_{i+1,j})\,\bigr\}, \label{eq:constraints:def}\\
 		& \;\; \vdots\nonumber\\
\Omega_{K} & =  \Omega_{IJ} =  \bigl\{\,\pb \mid\; \max(p_{I-1 J},p_{I,J-1}) &&< p_{IJ} &&< 1\,\bigr\}\nonumber.
\end{alignat}
where $K = IJ$. Note that we are using the bijection $\eta : [I] \times [J] \to [IJ]$ given by $\eta(i,j) = (i-1)J + j$ to transform a two-dimensional indexing to a one-dimensional index. For example, when $J= 4$ we have $\Omega_{5} = \Omega_{2,1}$ and so on. We will use these two indexings interchangeably throughout the paper. In particular, we often write the elements of $\pb$ in the one-dimensional index as well $\pb = (p_1,p_2,\dots,p_K)$.

The partial ordering constraints can be summarized as requiring 
\begin{align*}
	\pb \in \Omega, \quad \text{where} \quad \Omega := \bigcap^{K}_{k=1} \Omega_k \subset [0,1]^K.
\end{align*}
In the sequel, we refer to $\Omega$ as the lattice. We note that there is redundancy in the specification of $\Omega_k$'s in~\eqref{eq:constraints:def} in that the same constraint might be enforced by multiple $\Omega_k$'s. This redundancy is helpful in deriving the Gibbs sampler of Section~\ref{sub:gibbs}. 


\section{Nonparametric Bayesian model for dual agents}\label{sec:model}

We start by specifying our prior on $\pb$ and then discuss how we can sample from the prior and the posterior.

\subsection{Nonparametric prior}\label{subsec:npprior}
Perhaps the most basic nonparametric prior on $\pb$ is the uniform distribution on $\Omega$. 
%
%
The uniform distribution on the lattice has density
\begin{align}\label{eq:unif:f}
f_u(\pb) \propto 1_{\Omega}(\pb), \quad \text{where} \quad 1_{\Omega}(\pb) = \prod^{K}_{k=1} 1_{\Omega_k}(\pb)
\end{align}
is the indicator of the lattice. By the uniform distribution being nonparametric, we mean that one is not assuming a specific functional form for $p_{ij}$ based on a lower-dimensional parameter.
We can extend~\eqref{eq:unif:f} to a model with more general marginals. 
Assume that we want a prior on $\pb$ that is obtained as follows: Draw the coordinates of $\pb$ independently with $p_{ij}$ having density $b_{ij}(p_{ij})$, then truncate the joint distribution of $\pb$ to the set $\Omega$.
The density of this prior is given by
\begin{align}\label{eq:beta:f}
f(\pb) \propto 1_{\Omega}(\pb) \prod_{i,j} b_{ij}(p_{ij}).
\end{align}
In this paper, we take $b_{ij}(\cdot)$ to be beta densities:
\begin{align}\label{eq:bij:def}
b_{ij}(p_{ij}) \propto p_{ij}^{\alpha_{ij}-1} (1-p_{ij})^{\beta_{ij}-1}.
\end{align}
Note that we can write~\eqref{eq:beta:f} as
\begin{align}\label{eq:prior}
f(\pb) \propto \prod_{k=1}^K 1_{\Omega_k}(\pb) \,\prod_{i,j} b_{ij}(p_{ij}) 
\end{align}
which is a form suitable for Gibbs sampling since the lattice structure  encodes local relations between elements of $\pb$. Given the neighbors of a node in the lattice, its distribution is independent of the rest of the variables. In other words, $f(\pb)$ is a graphical model \citep{Koller:2009:PGM:1795555} with the undirected lattice diagram serving as its independence graph. For future reference, we will make the following definition:
\begin{defn}\label{defn:lat:res:Beta}
	 The \emph{lattice-restricted beta} distribution with shape parameters $\alphab=(\alpha_{ij})$ and $\betab=(\beta_{ij})$ is the multivariate distribution defined by~\eqref{eq:bij:def} and~\eqref{eq:prior}. We refer to $\alpha_{ij} +\beta_{ij}$ as the (effective) \emph{sample sizes} of the distribution (ESS).
\end{defn}

It is worth noting that the marginals of $\pb$ under a lattice-restricted beta distribution are not beta distributions themselves, due to the restrictions imposed by the lattice constraint. The notion of the sample size in Definition~\ref{defn:lat:res:Beta} is based on the common practice of referring to $\alpha + \beta$ as the effective sample size of beta$(\alpha,\beta)$ distribution. The rationale behind this naming is well-known from the posterior inference in beta-binomial models; see also~\eqref{eq:post:mean:beta:binom}. In a simple beta$(\alpha,\beta)$, the effective sample size $\alpha + \beta$ can directly control the variance of the distribution. For the multivariate   \emph{lattice-restricted beta}, the relation between the sample sizes $\alpha_{ij} + \beta_{ij}$ and the variances of the components of $\pb$ are much more complicated. In fact, the lattice constraint indirectly restricts how much $\alpha_{ij} + \beta_{ij}$ influences the variance, making it challenging to design diffuse priors. In Section~\ref{subsec:discount}, we propose a simple discounting scheme to work around this issue.

\subsection{Gibbs sampler for the prior}\label{sub:gibbs}
It is easy to sample from the lattice-restricted distribution~\eqref{eq:prior} using a Gibbs sampler. The updates are as follows:
Let us derive the updates for the Gibbs sampler: 
\begin{align}\label{eq:gibbs}
f(p_{11} \mid \pb_{-11}) &\;\propto\; b_{11}(p_{11}) \cdot 1_{\Omega_{11}}(\pb)\nonumber\\
&\;\;\;\vdots\nonumber\\
f(p_{ij} \mid \pb_{-ij}) & \;\propto\; b_{ij}(p_{ij}) \cdot 1_{\Omega_{ij}}(\pb) \\
&\;\;\;\vdots\nonumber\\
f(p_{IJ} \mid \pb_{-IJ}) & \;\propto\; b_{IJ}(p_{IJ}) \cdot 1_{\Omega_{IJ}}(\pb) \nonumber
\end{align}
where $\pb_{-ij}$ 
is the vector $\pb$ with $p_{ij}$ removed (i.e., all variables are included except $p_{ij}$). Note that although each $p_{ij}$ also appears in some other constraint sets besides $\Omega_{ij}$, we do not need to include them in the above conditional calculation since those constraints are also enforced by $\Omega_{ij}$. In other words, there is some redundancy in the condition of $\Omega_1, \Omega_2\dots, \Omega_K$ that we have introduced to simplify deriving the Gibbs sampler.

Each conditional distribution in~\eqref{eq:gibbs} is a truncated beta distribution which is easy to sample from, where the truncated beta density is defined as
\begin{align}
T(x;\alpha,\beta,a,b) \;\propto\; x^{\alpha-1} (1-x)^{\beta-1} 1\{x \in (a,b)\}.
\end{align}
Using this notation, for example,
\[
f(p_{11} \mid \pb_{-11})  = T\big(p_{11}; \, \alpha_{11},\beta_{11}, 0, \min(p_{12},p_{21}) \big).
\]
Thus, all these conditional distributions are easily derived and they are all truncated beta distributions that can be simulated efficiently. 
\subsection{Posterior}
Given prior~\eqref{eq:prior} and the likelihood in~\eqref{eq:likelihood}, we can readily obtain the posterior, 
\begin{equation}
\begin{aligned}
\pi(\pb \mid \zb) &\propto L(\zb \mid \pb) \;f(\pb) \\
&\propto \prod_{i,j} p_{ij}^{z_{ij}}(1-p_{ij})^{n_{ij} - z_{ij}} \prod_{k=1}^K 1_{\Omega_k}(\pb) \,\prod_{i,j} b_{ij}(p_{ij}) \\
&\propto \prod_{i,j} p_{ij}^{\alpha_{ij} + z_{ij}-1}(1-p_{ij})^{\beta_{ij} + n_{ij} - z_{ij}-1} \prod_{k=1}^K 1_{\Omega_k}(\pb)
\end{aligned}
\end{equation}
using~\eqref{eq:bij:def}. 
We note that the posterior is of the form
\begin{align}\label{eq:post}
\pi(\pb \mid \zb) &\propto \prod_{i,j} b'_{ij}(p_{ij}) \prod_{k=1}^K 1_{\Omega_k}(\pb)
\end{align}
where $b'_{ij}(\cdot)$ is the density of beta distribution with parameters $\alpha_{ij} + z_{ij}$ and ${\beta_{ij} + n_{ij} - z_{ij}}$. That is, posterior~\eqref{eq:post} is of the exact same form as~\eqref{eq:prior}, that is, a lattice-restricted beta distribution, with updated parameters. 
Thus, the Gibbs sampler derived earlier for the prior, works for the posterior as well, using the new beta parameters.

\subsection{Discounting the prior}\label{subsec:discount}
The relatively high-dimensional prior in~\eqref{eq:prior} will have reduced variances for components of $\pb$, relative to those, one would expect when the components are independent. This is due to the restrictions imposed by the lattice constraints and is a challenging aspect of specifying priors in high dimensions under many constraints on the coordinates. We believe the difficulty is present as long as one insists on the coordinates satisfying strict order constraints and is not an artifact of the particular choice of the beta densities.

At the early stages of the trial, due to the data having a small sample size and the prior having diminished variances (hence high concentration), the posterior inference will be dominated by the prior. 
This can be mitigated by controlling the sample size of the data relative to the (effective) sample size of the prior. To do so, we evaluate a pseudo-posterior by raising the likelihood to power $\omega > 1$ as follows:
\begin{equation}
\begin{aligned}
\pi(\pb \mid \zb) &\propto \Big(\prod_{ij} p_{ij}^{z_{ij}}(1-p_{ij})^{n_{ij}-z_{ij}} \Big)^\omega \prod_{ij} p_{ij}^{\alpha_{ij}-1} (1-p_{ij})^{\beta_{ij}-1} 1_\Omega(\pb) \\
&=\prod_{ij} p_i^{\omega z_{ij} + \alpha_{ij}-1} (1-p_{ij})^{\omega(n_{ij}-z_{ij}) + \beta_{ij}-1} 1_\Omega(\pb).
\end{aligned}
\end{equation}
The resulting pseudo-posterior is again an instance of a latticed-restricted beta distribution, as in Definition~\ref{defn:lat:res:Beta}, with shape parameters $\omega \zb +\alphab$ and $\omega(\nb-\zb) + \betab$.

The idea of raising the likelihood to a power has been explored in the literature to address model misspecification~\citep{royall2003interpreting,grunwald2017inconsistency,bissiri2016general} and to incorporate historical data in a Bayesian analysis~\citep{ibrahim2000power}. Using this idea to simulate diffuse priors from high-dimensional concentrated priors, as we intend here, is new to the best of our knowledge.
It is a natural approach for tuning the relative effects of the data and the prior on the posterior as the following simple example illustrates.

Consider the simple univariate model where $z \mid p \sim \text{Bin}(n,p)$  and $p \sim \text{beta}(\alpha, \beta)$. The \emph{$\omega$-reweighed pseudo-posterior} is a beta distribution with parameters $\omega z + \alpha$ and $\omega (n-z) + \beta$, whose mean is given by
\begin{align}\label{eq:post:mean:beta:binom}
\E[ p \mid z] &= \frac{\omega z + \alpha}{\omega n + \alpha +\beta} 
=: \lambda \frac{z}{n} +(1- \lambda) \frac{\alpha}{\alpha+\beta}
\end{align}
where $\lambda = \omega n / (\omega n + \alpha + \beta) \in (0,1)$. (Note that  in~\eqref{eq:post:mean:beta:binom}, $\alpha+\beta$ plays the same role in the prior term as does the sample size $n$ in the data-driven term.) That is, the posterior mean is a weighted (in fact, convex) combination of the maximum likelihood estimate $z/n$, which is solely based on data, and the prior mean,  with weights that are controlled by $\lambda$. Parameter $\omega$ allows us a degree of freedom beyond the sample size $n$ to control the effect of the prior, effectively tuning its overall variance. In particular, the weight of the data relative to the prior is given by
$\rho:= \frac{\lambda}{1-\lambda} = \frac{n \omega}{\alpha + \beta}$. For a desired level of $\rho$, which can be thought of as a user-specified level of confidence in the prior, we can solve for the appropriate $\omega$ as
\begin{equation}\label{eq:omega:rho}
\omega = \frac{\rho(\alpha + \beta) }{n}.
\end{equation}

By choosing $\rho$, one can calibrates the relative influence of the prior and data on the posterior.
When $\rho =1$, the relative influence of the prior and the data are as given by the traditional Bayesian approaches. 
For a more  outcome-adaptive inference, one sets $\rho$ to be greater than 1.  

 A value of $\rho > 1$ is what we suggest for the high-dimensional prior we are using (Definition~\ref{defn:lat:res:Beta}). As discussed earlier, the lattice constraint causes any prior distribution to have diminished variances. A choice of $\rho > 1$ deflates the effect of the prior, in effect simulating a more diffuse overall prior (i.e., having larger variance).
%
%
Empirically, we have found that setting $\rho=2$ significantly improves the performance. Thus, we choose
%
%
%
%
$\omega = {1+ 2\sum_{ij} (\alpha_{ij}+\beta_{ij})/\sum_{ij} n_{ij}}$
as suggested by~\eqref{eq:omega:rho} and since we need $\omega > 1$.

\subsection{Choosing the hyperparameters}\label{subsec:hyperparameter}
Let $m_{ij}$ be the effective sample size of the beta prior $p_{ij}$ (Definition~\ref{defn:lat:res:Beta}).
To choose the hyperparameters, we do a grid search over different choices of values for $m_{11}$ and $m_{IJ}$ with the the rest of $m_{ij}$s being equal to a value that is less than $\min(m_{11},m_{IJ})$. If the dose space is too large, one can limit the search space even further and assume that $m_{11} = m_{IJ}$. The grid search is done by running our Gibbs sampler algorithm \eqref{eq:gibbs} many times with all these different combination of hyperparameters. 
At the end, we choose the hyperparameters that match our prior guess of the toxicity probabilities.
The details for hyperparameter selection can be found in the Appendix \ref{sec:appendix_hyperparam}.

\section{Trial Design}\label{sec:trialdesign}
To limit the exposure of patients to toxic combinations and provide better posterior estimation, we enroll more patients to the first two cohorts. For better exploration of the dose space, following \cite{tighiouart2017bayesian}, we enroll patients to different dose combinations in each cohort $c > 1$. However, rather than alternating each time between the vertical and horizontal direction, we choose the direction randomly.

Thus, the design of a Phase I trial for two agents using the proposed NBCD proceeds as follows: 
\begin{enumerate}[label=(\arabic*)]
	\item  The first 4 patients in the first cohort receive the minimum dose combination $(d_1^A,d_1^B)$
	\item  In the second cohort, patients 5 and 6 receive $(d_1^A, d_{j^*}^B)$,
	where 
	\begin{align}\label{eq:hcrm}
		j^* = \argmin_j |\hat{p}_{1j} - \theta|
	\end{align}
     Similarly, patient 7 and 8 receive $(d_{i^*}^A, d_1^B)$,
     where 
     \begin{align}\label{eq:vcrm}
     i^*= \argmin_i |\hat{p}_{i1} - \theta|
     \end{align}
     \item In the $c$-th cohort ($c\geq 3$) of two patients, from each of the two dose combinations in cohort $c-1$, choose  between horizontal and vertical direction randomly to fix one drug level and vary the other drug level and find the dose combination with posterior median probability of DLT closest to $\theta$ similar to~\eqref{eq:hcrm} and~\eqref{eq:vcrm}. If the posterior median DLT probability of minimum dose of one direction or both directions is greater than $1.5\times \theta$, choose the direction with the lowest minimum one.  
     \item Repeat step (3) and terminate the trial when all the patients are enrolled, or the following stopping rule holds.	    
\end{enumerate}
 {\bf{Stopping rule:}}  Stop the trial after $n$ patients are accrued if
\begin{equation}
\pr(\hat{p}^{(n)}_{11} > \theta + \gamma) > \epsilon
\end{equation}
where, $\gamma$ and $\epsilon$ are the stopping rules parameters.  

\subsection{Recommended phase \rom{2} doses}
At the end of the trial, we recommend one or more dose combinations to be used in future phase \rom{2} studies. To achieve this, we first set the margins $\delta_l$ and $\delta_u$ and consider an asymmetric neighborhood $\mathcal{N}$ around $\theta$, that is $\mathcal{N} = [\theta - l, \theta -u] \subseteq [\theta - \delta_l , \theta - \delta_u]$. We start with small $l$ and $u$  and gradually increase them until for some $(i,j), \hat{p}_{ij} \in \mathcal{N}$. Among these dose levels, we recommend the ones that were experimented more than once (If no dose levels are available as such, we recommend the ones that are experimented once). If no dose levels belong to $[\theta - \delta_l , \theta - \delta_u]$, then we do not recommend any doses. Algorithm \ref{alg:doserecom} summarizes the steps for recommending doses for phase \rom{2}.

\begin{algorithm}[H]\label{alg:1}
	\caption{Dose recommendation for phase \rom{2}}
	\label{MTDalgorithm}
	\begin{algorithmic}[1]
	    \State Set $\delta_l$, $\delta_u$, the step sizes $\gamma_u$, $\gamma_l$, $\eta_u$ and pick initial values $l_0$, $u_0$.
	    \State Set $\mathcal{I} \leftarrow \varnothing; \quad l \leftarrow l_0; \quad u \leftarrow u_0$
		\If {\Big\{$\frac{\sum_{ij} 1\{\hat{p}_{ij} >  \theta\}}{K} \geq \frac{1}{2}$\Big\}}  
		\State The scenario is toxic; Set toxic $\leftarrow 1$.
		\EndIf         
		\While {$\mathcal{I} = \varnothing  \quad \text{AND} \quad (l \leq \delta_l \quad \text{OR} \quad u \leq \delta_u)$}
		\State Update $\mathcal{I} \leftarrow \Big \{(i,j): -l \leq \hat{p}_{ij} - \theta \leq u  \Big\}$
		\State Update $l \leftarrow l + 1\{l \leq \delta_l\} \times \gamma_l$
		\State Update $u \leftarrow u + 1\{u \leq \delta_u\} \Big(\eta_u \times \text{toxic} + \gamma_u \times (1-\text{toxic})\Big)$
		\EndWhile
		\State Return $\widehat{\text{MTD}}(\delta) = \Big\{ d_{ij} :  (i,j) \in \mathcal{I} , n_{ij} > 1\Big\}$ 
	\end{algorithmic}
	\label{alg:doserecom}
\end{algorithm}


\section{Simulations}\label{sec:simulations}
In this section, we show the effectiveness of our proposed method in comparison with the existing methods through various simulation studies.  
For all the following simulations and the real trial design, we define the MTD($\delta$) as any dose combination that is within $\delta = 0.1$ of the the target probability and the MTD($\delta$) is estimated using Algorithm~\ref{MTDalgorithm}. We also use $\delta_l = 0.1$ and $\delta_u = 0.05$, $l_0 = 0.05$, $u_0 = 0$, $\gamma_l = \frac{\delta_l}{2}$, $\gamma_u = \frac{\delta_u}{2}$ and $\eta_u = \frac{\delta_u}{5}$, throughout.
All the trials start from the lowest dose level $d_{11} = (d_1^A, d_1^B)$. We use a cohort size of 4 for the first two cohorts and a cohort size of 2 for the rest. The toxicity outcome is generated as a Bernoulli random variable that takes a value of 1 with probability $p_{ij}$ and 0 from the corresponding scenario. 
For finding the median posterior, we took 11000 posterior samples and discarded 1000 burn-in iterations in Gibbs sampling procedure. To select the hyperparameters, we used a grid search as explained in section~\ref{subsec:hyperparameter}.
For each scenario, 2000 simulated trials are replicated to evaluate the operating characteristics of NBCD and other methods. Specifically, we calculated the mean percentage that each dose combination was selected as the MTD($\delta$) at the end of the trial (recommendation percentage) and the mean percentage of patients assigned to each dose combination (experimentation percentage). 
For PIPE design, the (weak) prior sample size of $\frac{1}{I*J}$ was used as suggested by \cite{mander2015product}. The dose escalation is done by a neighborhood constraint, with admissible doses chosen from those closest to the estimated MTD contour. 




\subsection{Simulation Study \rom{1}}\label{subsec:sim1}
For the first simulation study, we compare the performance of NBCD with the results from \cite{mander2015product} (previously examined by \cite{braun2013generalized}) under scenarios A-G that are reproduced in Table~\ref{tab:braunscenarios}. These methods include the generalized CRM (gCRM) \cite{braun2013generalized}, the coupla model of \cite{yin2009Bayesian} and the latent contingency method by \cite{yin2009latent} and the product of independent beta probabilities (PIPE) \cite{mander2015product}.  Among all these methods, PIPE and NBCD are nonparametric and the rest are parametric models. 

\begin{table}
\caption{\label{tab:braunscenarios}Dose limiting toxicity for simulation study \rom{1}}
\begin{tabular}{*{14}{c}}
&    &      & & &   & Drug A & & &&  & &&\\
&     &  1 & 2 & 3 & 4 &&&1&2&3&4&& \\
\hline\\
&&\multicolumn{4}{c}{Scenario A}  & &&\multicolumn{4}{c}{Scenario E} && \\
\rule{0pt}{3ex} 
	 & 1 & 0.04  & 0.08 & 0.12 &  0.16 &  &  & 0.08 & 0.18 & 0.28 &  0.29 &&\\
	 & 2 & 0.10 & 0.14  & 0.18  & 0.22 & &  & 0.09 & 0.19& 0.29& 0.30 &&\\
	 & 3 & 0.16 & 0.20 & 0.24 & 0.28 & & & 0.10 & 0.20 & 0.30 & 0.31 &&\\
	 & 4 & 0.22 & 0.26 & 0.30 & 0.34 & &  & 0.11 & 0.21 & 0.31 & 0.41 &&\\
	 \rule{0pt}{4ex} 
	&&\multicolumn{4}{c}{Scenario B}  & &&\multicolumn{4}{c}{Scenario F}  &&\\
\rule{0pt}{3ex} 
	  & 1 & 0.02 & 0.04 & 0.06 & 0.08 & &  & 0.12 & 0.13 & 0.14 & 0.15 &&\\
	    & 2 & 0.05 & 0.07 & 0.09 & 0.11 & & & 0.16 & 0.18 & 0.20 & 0.22 &&\\
	    & 3 & 0.08 & 0.10 & 0.12 & 0.14 & &  & 0.44 & 0.45 & 0.46 & 0.47&&\\
	    & 4 & 0.11 & 0.13 & 0.15 & 0.17 & &  & 0.50 & 0.52 & 0.54 & 0.55 && \\
	    \raisebox{-.5\normalbaselineskip}[0pt][0pt]{\rotatebox[origin=c]{90}{Drug B}} &  & & &  & & &  & & & &&\\
	    &&\multicolumn{4}{c}{Scenario C}  & &&\multicolumn{4}{c}{Scenario G} && \\
\rule{0pt}{3ex} 
	     & 1 & 0.10 & 0.20 & 0.30 & 0.40 &  &  & 0.01 & 0.02 & 0.03 & 0.04 &&\\
	    & 2 & 0.25 & 0.35 & 0.45 & 0.55 & & & 0.04 & 0.10 & 0.15 & 0.20 &&\\
	    & 3 & 0.40 & 0.50 &0.60 & 0.70 & &  & 0.06 & 0.15 & 0.30 & 0.45 &&\\
	    & 4 & 0.55 & 0.65 & 0.75 & 0.85 & &  & 0.10 & 0.30 & 0.50 & 0.80 && \\
	     \rule{0pt}{4ex} 
	    &&\multicolumn{4}{c}{Scenario D}  &&&&&& &&\\
	  \rule{0pt}{3ex} 
	     & 1 & 0.44 & 0.48 & 0.52 & 0.56 & & & & & & & &\\
	    & 2 & 0.50 & 0.54 & 0.58 & 0.62 & & & & & &  &&\\
	    & 3 & 0.56 & 0.60 & 0.64 & 0.68 & & & & & &  &&\\
	    & 4 & 0.62 & 0.66 & 0.70 & 0.74 & & & & & &  &&\\
	    \hline
\end{tabular}
\end{table}

The target toxicity probability $\theta$ is set at 0.2 and the total sample size is 50. 
The median prior for the probability of DLT at the smallest and largest dose combinations is set to 0.04 and 0.34, respectively to match that of scenario A. The stopping rules parameters $\gamma = 0.1$ and $\epsilon = 0.8$ are used.  

\begin{table}
\caption{\label{tab:sim1result}Experimentation and recommendation percentages for simulation study \rom{1}}
	\footnotesize
	\begin{tabular}{cc|cccc|cccc}
		 &&\multicolumn{4}{c}{Recommendation percentages}  &\multicolumn{4}{c}{Experimentation  percentages}\\
		 \hline
		 \rule{0pt}{3ex} 
		&  &  At $\theta$ &  1-10\% & $> 10 \%$ & None &  At $\theta$ &  1-10\% & $>10\%$  & None\\
		Scenario &  Model  &  & of $\theta$  &   of $\theta$ &  &  &         of $\theta$  &  of $\theta$ &  \\
		\hline
		 \rule{0pt}{3.5ex} 
		A &  gCRM  & 10 & 82 & 3 & 5 & 6 & 72 & 17 & 5\\
		    & YY09a & 13 & 82 & 5 & 0 & 13 & 72 & 15 & 0\\
		    & YY09b & 11 & 81 & 6 & 2 & 10 & 70 & 20 & 0\\
		    & PIPE  & 10 & 88 & 3 & 0 & 8 & 87 & 5 & 0\\
		    & NBCD   & 16 &  83  &  1    &  0 & 10  &  76   &  14  &  0    \\
		     \rule{0pt}{3.5ex} 
		B &  gCRM  & 0 & 94 & 3 & 3 & 0 & 87 & 13 & 0\\
		   & YY09a & 0 & 99 & 1 & 0 & 0 & 86 & 14 & 0\\
		   & YY09b & 0 & 96 & 4 & 0 & 0 & 71 & 29 & 0\\
		   & PIPE  & 0 & 83 & 17 & 0 & 0 & 82 & 18 & 0\\
		   &  NBCD   & 0 & 96    &  4    &   0 &  0   &  72  & 28   & 0  \\  
		    \rule{0pt}{3.5ex} 
		  C &  gCRM  & 45 & 39 & 5 & 11 & 30 & 41 & 18 & 11\\
		     & YY09a  & 41  & 50 & 5 & 4 &  27 & 54 & 16 & 3\\
		     & YY09b & 42 & 47 & 5 & 6 & 29 & 55 & 11 & 5\\
		     & PIPE  & 29 & 59 & 7 & 5 & 19 & 46 & 34 & 2\\
		     & NBCD  & 32  &  54   &   14  &  0  &  20  &  48  &  31   &  1    \\
		  \rule{0pt}{3.5ex}   
		   D &  gCRM  & 0 & 0 & 4 & 96 & 0 & 0 & 22 & 78\\
		  & YY09a  &0 & 0 & 1 & 99 & 0 & 0 & 20 & 80\\
		  & YY09b & 0 & 0 & 1 & 99 & 0 & 0 & 16 & 84\\
		  & PIPE  & 0 & 0 & 1 & 99 & 0 & 0 & 37 & 63\\
		  & NBCD    & 0     &  0     &  4  & 96  &    0 &    0 &   41 & 59 \\
		  \rule{0pt}{3.5ex} 
		  E&  gCRM  & 9 & 70 & 14 & 7 & 5 & 56 & 32 & 7\\
		  & YY09a   & 6 & 65 & 27 & 2 & 9 & 55 & 34 & 2\\
		  & YY09b   & 7 & 67 & 25 & 1 & 6 & 54 & 38 & 2\\
		  & PIPE      & 11  & 84& 4 & 1 &  9& 77 & 13 & 1\\
		  & NBCD    &  15  & 78     &   7 &    0&   10  &  70   & 20    & 0 \\  
		   \rule{0pt}{3.5ex} 
		   F&  gCRM  & 13 & 70 & 6 & 11 & 10 & 64 & 16 & 10\\
		   & YY09a   & 14 & 76 & 6 &  4&  7 & 75 & 14 & 4\\
		   & YY09b   & 12 & 74 & 7 & 7 &  7& 77 & 9 & 7\\
		   & PIPE      & 12  & 75& 11 & 2 &  12& 69 & 18 & 2\\
		   & NBCD    &  16  &  72  &  11 & 1  &  14  &   65  &   20 & 1 \\  
		    \rule{0pt}{3.5ex} 
		   G&  gCRM  & 25 & 68 & 5 & 2 & 18 & 57 & 24 & 1\\
		   & YY09a   & 12 & 76& 12 &  0&  3  & 71 & 26 & 0\\
		   & YY09b   & 15 & 72 & 13 & 0 &  7 & 61 & 32 & 0\\
		   & PIPE      & 9  & 62& 29 & 0 &  14 & 54 & 31 & 0\\
		   & NBCD     & 20  &  74  & 6 & 0&   15  &  56   &  29 & 0  \\  		      
	\end{tabular}
\end{table}
The operating characteristics of the NBCD method and all the other methods are shown in Table~\ref{tab:sim1result}, where the results from the parametric models and the PIPE method were produced from Table \rom{4} of \cite{mander2015product}.
In scenario D where all doses are toxic, all methods perform well in the sense that they do not recommend an MTD. Our method outperforms all the other methods in scenarios A,E,F in dosing at the target and in scenarios A,F,G in dose recommendation within 10\% of the target. In particular, the percent recommendation within 10\% of the target for NBCD exceeds that of PIPE by an absolute 13\% and 23\% under scenarios A and G, respectively. 
The percent of patients allocated to doses within 10\% of the target is higher for PIPE relative to NBCD under scenarios A, B, and E and they are fairly close for scenarios C, F, and G. Given that the primary goal of phase I trials is estimation of the MTD, we conclude that our method is competitive with the other approaches under the scenarios studied by \cite{braun2013generalized, mander2015product}.
\subsection{Simulation Study \rom{2}}\label{subsec:sim2}
For our second simulation study, we investigate the performance of our method under an asymmetric dose-combination space with seven 4 by 5 scenarios, see Table \ref{tab:4by5scenarios}.
\begin{table}
\caption{\label{tab:4by5scenarios}Dose limiting toxicity scenarios in simulation study \rom{2}}
	\begin{tabular}{*{14}{c}}
		\hline\\
		&    &      & & &   && Drug A & & &&  & &\\
		&     &  1 & 2 & 3 & 4 & 5&&&1&2&3&4 & 5\\
		\hline\\
		&&\multicolumn{5}{c}{Scenario 1}  & &&\multicolumn{5}{c}{Scenario 2}  \\
		\rule{0pt}{3ex} 
		& 1 & 0.05 & 0.07 & 0.11 & 0.16 & 0.23   &&& 0.01 & 0.03 & 0.07& 0.09 & 0.11\\
		& 2 & 0.07 & 0.12 & 0.17 &  0.24 & 0.33  &&& 0.04 & 0.06 & 0.08 & 0.10& 0.22 \\
		& 3 & 0.12 & 0.18 & 0.25 & 0.33 & 0.43   &&& 0.09 & 0.13 &  0.22& 0.25& 0.27\\
		& 4 & 0.18 & 0.27 & 0.35 & 0.43 & 0.50   &&& 0.12 & 0.16 & 0.23 &  0.28 & 0.30\\
		\rule{0pt}{3ex} 
		&&\multicolumn{5}{c}{Scenario 3}  & &&\multicolumn{5}{c}{Scenario 4}  \\
		\rule{0pt}{3ex} 
		& 1 & 0.30 & 0.35 &  0.40 & 0.50 & 0.55 & && 0.01 & 0.03 & 0.08 & 0.12 & 0.15\\
			 \raisebox{-.5\normalbaselineskip}[0pt][0pt]{\rotatebox[origin=c]{90}{Drug B}}& 2 & 0.40 & 0.55 &  0.65 & 0.75&  0.85 &&& 0.02 & 0.05 & 0.10 & 0.16 & 0.30\\
	    & 3 & 0.50 & 0.60 & 0.70 & 0.80 &  0.90 &&& 0.07 & 0.09 & 0.15 & 0.25 & 0.35 \\
		& 4 & 0.55 & 0.70 &  0.75 & 0.85 & 0.95 &&& 0.10 & 0.26 & 0.30 & 0.33 & 0.50\\
		\rule{0pt}{3ex} 
		&&\multicolumn{5}{c}{Scenario 5}  & &&\multicolumn{5}{c}{Scenario 6}  \\
	
		\rule{0pt}{3ex} 
        & 1 & 0.07 & 0.12 & 0.20 & 0.25 & 0.30 &&& 0.10 &  0.15 &  0.20 & 0.30 & 0.45\\
		& 2 & 0.10 & 0.18 & 0.23 & 0.30 & 0.35 &&& 0.11 & 0.20 & 0.30 & 0.40 & 0.50  \\
		& 3 & 0.30 & 0.48& 0.56 & 0.65 & 0.68 &&& 0.15 & 0.30 &  0.35 &  0.50 & 0.60\\
		& 4 & 0.40 & 0.55 & 0.60 & 0.66 & 0.70 &&& 0.30 & 0.40 & 0.50 & 0.60 & 0.65 \\

		&&\multicolumn{5}{c}{Scenario 7}  \\
		\rule{0pt}{3ex}
		& 1 & 0.11 & 0.12 & 0.13 & 0.14 & 0.15  & & &  &  &  &  & \\
		& 2 & 0.14 & 0.20 & 0.25 & 0.30 & 0.35 &&&   & &  &  & \\
		& 3 & 0.16 & 0.25 & 0.40 & 0.55 & 0.60 &&& & & &  &  \\
		& 4 & 0.20 & 0.40 & 0.60 & 0.90 & 0.95 &&& & & &  & \\
		\hline
	\end{tabular}
\end{table}
These scenarios cover a wide range of dose–response relationships and include cases where the MTD is achieved at the highest dose combination (scenario 2), lowest dose combination (scenario 3), and more complex structures (scenarios 5 and 7). We compare our method with PIPE as it is a nonparametric model. The target toxicity probability $\theta$ is set at 0.3 and the total sample size is 50 with a cohort size of 2 for PIPE. For a fair comparison, we do not impose any early termination for NBCD and PIPE.
The median prior for the probability of DLTs at the smallest and largest dose combinations was set to 0.05 and 0.50, respectively. We then chose the hyperparameters that match our prior guess of the toxicity probabilities as discussed in section~\ref{subsec:hyperparameter}.
We used scenario 1 as the prior for both NBCD and PIPE. Table~\ref{tab:sim2result} shows that NBCD outperforms PIPE in all scenarios with respect to percent recommendation within 10\% of the target probability of DLT $\theta$ with the highest percent equal to 20\% achieved in scenarios 2 and 3. 
\begin{table}
\caption{\label{tab:sim2result}Experimentation and recommendation percentages for simulation study \rom{2}}
	\small
	\begin{tabular}{cc|cccc|ccc}
		&&\multicolumn{4}{c}{Recommendation \%}  &\multicolumn{3}{c}{Experimentation \%}\\
		\hline
		\rule{0pt}{3ex} 
		&  &  At $\theta$ &  1-10\% &  $>10\%$ & None &  At $\theta$ &  1-10\% & $>10\%$\\
		Scenario &  Model  &  & of $\theta$  &   of $\theta$ &  &  &  of $\theta$  &  of $\theta$\\
		\hline
		\rule{0pt}{4ex} 
		1& NBCD  & 0.0 & 84.0  & 16.0  & 0.0  & 0.0  & 58.1  & 41.9\\
		&  PIPE  & 0.0  & 75.0 & 25.0 & 0.0 &  0.0  & 60.9 & 39.1\\
		\rule{0pt}{4ex} 
		2& NBCD & 6.8 & 81.3 & 11.5 & 0.4   & 6.3 & 53.8 & 39.9\\
		 &  PIPE  & 4.2 & 64.3 & 31.5 & 0.0 & 5.5 & 56.8 & 37.7\\
		\rule{0pt}{4ex} 
		3 & NBCD  & 24.6 & 60.7 &  10.5 & 4.2  &  26.4 & 49.6 & 24.0\\
		 & PIPE  & 24.0 &  40.5&  35.4 & 0.1 & 19.4 & 33.3 & 47.3\\
		\rule{0pt}{4ex} 
		4& NBCD  & 29.7  & 52.5 & 17.5  & 0.3 & 20.4 &  36.6  & 43.0\\  
		 &  PIPE  & 25.4 & 38.4  & 36.2  & 0.0  & 19.4 & 32.4 & 48.2\\
		\rule{0pt}{4ex} 
		5& NBCD  & 36.7& 41.4 & 20.5 & 1.4  & 25.3  & 29.5& 45.2\\
		& PIPE  & 36.5 & 39.6 & 23.6 & 0.3 & 22.3 & 29.3 & 48.4\\
		\rule{0pt}{4ex} 
		6& NBCD & 52.3 & 38.8 & 8.6 & 0.3 & 35.3 & 34.1 & 30.6\\
		 &  PIPE  & 40.7 & 35.9 & 23.0 & 0.4 & 30.7 & 36.9& 32.4\\
		\rule{0pt}{4ex}
		7& NBCD  & 16.8 & 72.9  & 9.8  & 0.5  & 9.8 & 56.5  & 33.7\\ 
	     &  PIPE  & 10.0 &  60.6&  29.4& 0.0 &  10.2 & 52.9 & 36.9\\
		
	\end{tabular}
\end{table}
The percent of patients allocated to doses within 10\% of the target are fairly similar between the two methods except for scenario 3 where NBCD allocates 23\% more patients than PIPE. The safety profiles of the two methods are presented in Table~\ref{tab:sim2safety}. 
\begin{table}
\caption{\label{tab:sim2safety}Trial safety evaluation for simulation study \rom{2}
}
	\resizebox{\textwidth}{!}{%
	\begin{tabular}{lcccccccc}
	    Design & 1 & 2 & 3 & 4 & 5 & 6 & 7 & Average\\
		\hline
		 Recommendation \% of overtoxic doses& & & & & & &\\
		 NBCD  & 4.4 &  0.0* &  9.7   &  0.3 &  12.4  &  5.3 & 1.3 & 5.6\\
		 PIPE  & 5.6  &  0.0* &  22.4 &  0.0 &  10.5 &  13.7 & 2.8 & 9.2\\
		 \hline
		 Allocation \%  of patients to overtoxic doses & & & & & & &\\
		 NBCD  & 6.0 &  0.0* & 24.0   & 1.4  & 22.2  & 9.9  &  4.9  &  11.4\\
		 PIPE  & 8.7  & 0.0* &  40.4 &  2.8  & 29.2 &  19.4 &  11.5 & 18.7\\
		 \hline
		 Average rate of DLTs & & & & & & & \\
		 NBCD  & 23.3 &  18.4 &  39.6  &  20.8 &  30.2 & 29.3  & 26.7 & 26.9 \\
		 PIPE  & 25.8  & 20.2 &  43.7 &   22.8 &  34.2 & 32.6 & 29.5 & 29.8\\
		 \hline
		 \% of trials with DLT rate $> \theta + \delta$& & & & & & &\\
		 NBCD  & 0.0 & 0.0 & 34.7 & 0.0 & 1.6 &  0.1 & 0.0 & 5.2\\
		 PIPE     & 0.1 & 0.0 & 67.1 & 0.0 & 9.7 &  4.4 & 1.1 & 11.8  \\
		 \scriptsize{* The average excludes the items with asterisk}\\
	\end{tabular}
	}
\end{table}
In general, the average percent of DLTs across all simulated trials are fairly close in all seven scenarios but PIPE tends to allocate more patients to overtoxic doses and is more likely to recommend overtoxic doses under scenarios 3 and 6. Finally, PIPE is more likely to result in a trial with an excessive rate of DLTs relative to NBCD as assessed by the percent of trials with a DLT rate more than $\theta+0.1$. In particular, the probability that a prospective trial using PIPE will result in an excessive rate of DLTs under scenario 3 exceeds that of NBCD by 32.4\%. Based on these scenarios, we conclude that NBCD approach is safer than PIPE and more efficient in recommending the MTD.

\section{Trial Application}\label{sec:realtrial}
We used our approach NBCD to design a phase I trial of the combination Gemcitabine (Gem) and CB-839, an orally bioavailable inhibitor of Glutamine in advanced stage pancreatic cancer patients refractory to first-line FOLFIRINOX (5-Fluorouracil (5-FU), Leucovorin, Irinotecan, and Oxaliplatin) at our institution. 
We will explore two dose levels of Gem 800 and 1000 \si{mg/m^{2}}  and three dose levels of CB-839, 400, 600, and 800 \si{mg} BID for a total of 6 possible dose combinations and we plan to enroll $n = 36$ patients to the trial. The target probability of DLT is set to $\theta = 0.33$ and the median prior for the probability of DLTs at the smallest and largest dose combinations were set to 0.05 and 0.30, respectively. These restrictions were achieved after discussion with the clinician who used the fact that CB-839 was well tolerated at the highest dose when combined with other cytotoxic drugs from previous phase I/II trials. The stopping rules parameters $\gamma = 0.1$ and $\epsilon = 0.8$ were used. We evaluated the performance of NBCD and PIPE by simulating 2000 trials under the five scenarios shown in Table~\ref{tab:realtrial}.
\begin{table}
\caption{\label{tab:realtrial}Dose limiting toxicity scenarios in real trial design}
	\centering
	\begin{tabular}{*{11}{c}}
		\hline\\
		&&    &      & & &    CB-839 & & &&  \\
		& &    &  400 & 600 & 800 &&&400&600&800\\
		\hline\\
		&&&\multicolumn{3}{c}{Scenario 1}  & &&\multicolumn{3}{c}{Scenario 2}  \\
		\rule{0pt}{3ex} 
		& 800 && 0.05 &  0.10&  0.20    &&& 0.05 & 0.20 & 0.33 \\
		& 1000&& 0.10 & 0.20 & 0.30   & & & 0.15 & 0.33 & 0.50 \\
		\rule{0pt}{4ex} 
		&&&\multicolumn{3}{c}{Scenario 3}  & &&\multicolumn{3}{c}{Scenario 4}  \\
		\rotatebox[origin=c]{90}{Gem} & 800 && 0.01 & 0.03 &  0.07 & && 0.30 & 0.40 & 0.50\\
		& 1000 && 0.03 & 0.07 &  0.10 &&& 0.40 & 0.55 & 0.65 \\
		\rule{0pt}{4ex} 
		&&&\multicolumn{3}{c}{Scenario 5} & && \multicolumn{3}{c}{ } \\
		& 800 && 0.30 & 0.50 &  0.60 & &&  &  & \\
		& 1000 && 0.54 & 0.58 &  0.65 &&&  &  & \\
		\hline
	\end{tabular}
\end{table}

For PIPE, a cohort of size 2 was used.
Scenario 1 is a case where the MTD is achieved at dose combination 800 \si{mg} CB-839 and 1000 \si{mg/m^2} gemcitabine. In scenario 2, there are two MTDs. Scenario 3 has no MTDs. Scenario 4 has three MTDs and in scenario 5,  the MTD is achieved at the minimum dose combination. The operating characteristics are presented in Tables~\ref{tab:realtrialresult} and ~\ref{tab:safetyreal}. 
\begin{table}
\caption{\label{tab:realtrialresult}Experimentation and recommendation percentages for the real trial}
	\small
	\begin{tabular}{cc|cccc|cccc}
		&&\multicolumn{4}{c}{Recommendation \%}  &\multicolumn{4}{c}{Experimentation \%} \\
		\hline
		\rule{0pt}{3ex} 
		&  &  At $\theta$ &  1-10\% & $> 10 \%$ & None &  At $\theta$ &  1-10\% & $>10\%$  & None\\
		Scenario &  Model  &  & of $\theta$  &   of $\theta$ &  &  &         of $\theta$  &  of $\theta$ &  \\
		\hline
		\rule{0pt}{4ex} 
		1 & NBCD  & 0.0 & 66.8 & 27.8 & 5.4 & 0.0 & 51.4 & 48.6 & 0.0\\
		 &  PIPE  & 0.0 & 23.6 &  75.9 & 0.5  & 0.0 & 36.3  & 63.4 & 0.3 \\ 
		\rule{0pt}{4ex} 
		2 & NBCD & 70.4  & 0.0 &  25.7 & 3.9 & 45.6 & 0.0 & 54.4 & 0.0  \\
		 &  PIPE  & 47.8 & 0.0 &  51.3 & 0.9 & 53.4  & 0.0 & 46.4 & 0.2 \\	
		\rule{0pt}{4ex} 
		3& NBCD  & 0.0  & 0.0 & 16.8 & 83.2 &  0.0  & 0.0 & 100  & 0.0\\
		 & PIPE  & 0.0 & 0.0  & 100  & 0.0    & 0.0  & 0.0 & 100  & 0.0 \\	
		\rule{0pt}{4ex}
		4 & NBCD  & 0.0 &  81.6 & 6.3 & 12.1 & 0.0 & 76.5 & 16.4 & 7.1\\   
		  &  PIPE  & 0.0 & 72.2 & 5.3 & 22.5 & 0.0 & 67.4 & 21.9 & 10.7\\
		\rule{0pt}{4ex}  
		5 & NBCD  & 0.0 & 53.2 & 23.8 & 23.0 & 0.0 & 40.8 & 52.3 & 6.9\\   
		  &  PIPE  & 0.0 & 54.1 & 12.9 & 33.0& 0.0 & 41.5 & 46.4 & 12.1  
	\end{tabular}
\end{table}
The percent recommendation within 10\% of the target dose using our approach is significantly higher than that of PIPE under scenarios 1, 2, and 4 and the two approaches are equivalent under scenario 5. For Scenario 3, although the probability of DLT at the maximum dose combination is only 0.1, PIPE always recommends a dose. However, NBCD does not recommend any doses in 83.2\% of the trials. Furthermore, NBCD is safer as assessed by the percent of trials with an excessive rate of DLT under scenarios 4 and 5. Based on the high percent recommendation and safety summary statistics, we conclude that our approach achieves good operating characteristics under various scenarios for the location of the MTD.
\begin{table}
\caption{\label{tab:safetyreal}Trial safety evaluation for the real trial}
	\small
	\begin{tabular}{lcccccccc}
	    Design & 1 & 2 & 3 & 4 & 5  & Average\\
		\hline
		 Recommendation \%  of overtoxic doses & & & & &  \\
		 NBCD  & 0.0 & 10.2  & 0.0*  &  6.3    & 23.8  &  13.4 \\
		 PIPE  & 0.0 & 1.0  & 0.0*  & 5.3   &  12.9 & 6.4 \\
		 \hline
		 Allocation \%  of patients to overtoxic doses & & & & & \\
		 NBCD  & 0.0  & 22.3 & 0.0*  &  16.4   & 50.4 &  29.7 \\
		 PIPE  & 0.0 & 8.9 & 0.0* & 21.9 & 46.4  & 25.8  \\
		 \hline
		 Average rate of DLTs & & & & &  \\
		 NBCD  & 21.8  & 30.3  &  7.9  & 36.4 & 39.0 & 27.1  \\
		 PIPE  &  21.9 & 28.0 & 9.0 &  45.9  & 48.8 & 30.7 \\
		 \hline
		 \% of trials with DLT rate $> \theta + \delta$& & & & \\
		 NBCD    &  0.2  &  0.5 &   0.0 & 19.5 & 39.8 & 12.0\\
		 PIPE       &  0.2  &   0.3  & 0.0  &  32.6   & 49.5 & 16.5\\
		 \scriptsize{* The average excludes the items with asterisk}\\
	\end{tabular}
\end{table}
\section{Concluding Remarks}\label{sec:conclusion}
In this paper, we considered a nonparametric Bayesian approach to dose finding for combinations of drugs (NBCD) in phase \rom{1} clinical trials. This approach does not rely on a specific functional form of the dose-toxicity relationship and hence does not suffer from lack of robustness under strong model misspecification seen in parametric methods. Our model imposes the partial ordering constraints on the probabilities of toxicities by truncating the independent beta distributions to the set of partial orders. Using the fact that the conditional distributions of the probabilities of DLT given the remaining risk of toxicities are truncated beta densities, features of the joint posterior probability of DLT are easily estimated using the Gibbs sampler. This is in contrast to PIPE where independent prior beta distributions are placed on the space of dose combinations; the partial ordering constraint is only enforced on the estimated maximum tolerated contours during the trial. In addition, there is more borrowing of information across the dose combination space using NBCD relative to PIPE due to the partial ordering constraint {\it a priori}.

Due to the complexity of the joint prior probability of DLT $\pb$, we selected a sub-class of priors such that the hyperparameters $(\alpha_{ij},\beta_{ij})$ are all equal, except perhaps for the minimum and maximum dose combinations. These hyperparameters are then selected using a grid search algorithm described in Appendix A. In our simulation studies and the trial example, we ask the clinician to specify prior median probabilities of DLT at the minimum and maximum dose combinations and among all hyperparameters that satisfy this constraint, we select the combination that maximizes the trace of the covariance matrix of the prior joint probability of DLT $\pb$. This simplified sub-class of priors is not computationally intensive and performs well in our simulations under a large class of scenarios including cases proposed by other authors. In addition, it is practically appealing to clinicians since it is easier to specify {\it a priori} probabilities of DLT at the minimum and maximum dose combination than it is for all combinations.

Dose allocation to successive cohorts of patients follows a modified continual reassessment scheme. It allows for better exploration of the dose combination space, resulting in high recommendation percentages relative to previous approaches. It is an extension of the algorithms proposed in \cite{tighiouart2017bayesian}, \cite{diniz2017bayesian} and \cite{jimenez2019cancer} where escalation of the two drugs simultaneously is not allowed but the drug to be escalated (or de-escalated) is selected randomly. We have shown that NBCD has comparable operating characteristics with other parametric approaches (Simulation Study \rom{1} that is based on scenarios studied by \cite{braun2013generalized}) and outperforms PIPE in most scenarios for Simulation Study \rom{1}-\rom{2} and the real trial example both in terms of percent MTD recommendation and safety. We are currently working on extending this nonparametric design in the setting of phase I/II drug combination trials where a separate dose efficacy model is specified for combinations of cytotoxic and biologic drugs. While the extension using two cytotoxic agents is straight forward, using one or more biologic drugs is more challenging to model non-parametrically since efficacy may increase at low dose levels and then plateau past a threshold contour. 

In conclusion, NBCD is a good alternative approach to parametric model based designs when robustness to model misspecification is a concern and it still borrows information across the dose combination space owing to the partial ordering constraint on the prior probabilities of DLT. It outperforms the nonparametric approach PIPE in the simulation studies we presented and the real trial example. However, it is computationally more intensive than PIPE but operating characteristics for prospective studies can be carried out in a few hours. Computer R codes are available from the authors upon request.
\appendix
\section{Details for hyperparameter selection} \label{sec:appendix_hyperparam}
\appendix
We choose $\alphab$ and $\betab$ vectors to have the following forms:
\begin{align*}
	\begin{array}{clcccl}
	\alphab = &  \big[\, m(1-t) & l & \dots & l &  Ms \,\big],
	\\[1ex]
	\betab = & \big[\, mt & u & \dots & u &  M(1-s) \,\big].
	\end{array} 
\end{align*}
That is, all the elements of $\alphab$, except the first and last, are taken to be equal, and similarly for $\betab$. The values $m,M,s,t,l$ and $u$ are determined by a grid search as follows.

We vary $t$ and $s$ in a subinterval of $(0,1)$ away from the boundaries, e.g., $(0.2,0.5)$. We choose $u$ and $l$ such that $u+l \le \min(m,M)$. To reduce the size of the grid search, we can fix one of $u$ or $l$. For example, we can set $u = \min(m,M)/2 - l$ and let $l$ vary in the interval $[0.2,0.4]\times\min(m,M)$. The two parameters $m$ and $M$ are also varied, independently, over an interval $[\min(\alpha_0,\beta_0), \alpha_0 + \beta_0]$ where $\alpha_0$ and $\beta_0$ are some heuristic prespecified values. For example, $\alpha_0 + \beta_0$ roughly specify the overall sample size of the prior. Below we will discuss a heuristic for choosing $\alpha_0$ and $\beta_0$ that we have found effective in practice.

The goal of the gird search is to find a combination of hyperparameters such that the resulting prior satisfies some specified criteria for the median probability at selected dose combinations. Let $\tilde{p}_{ij}$ be the prior median for the $(i,j)$-th does combination. Often the median for the smallest and largest dose combinations, i.e., $\tilde{p}_{11}$ and $\tilde{p}_{IJ}$, are required to match certain values and we have a range for the intermediate dose combinations. For example, the following is a possible set of criteria for a $4\times 4$ lattice:
\begin{align*}
	\{ \tilde{p} \mid \; \tilde p_{11} \approx 0.04, \; \tilde p_{44} \approx 0.34, \; \tilde p_{12} < 0.1\}.
\end{align*}
This set is often rewritten by setting a tolerance, say $\delta=0.01$,
\begin{align}\label{eq:temp:7304}
\{ \tilde{p} \mid \; |\tilde p_{11} - 0.04 | < \delta , \; |\tilde p_{44} - 0.34| < \delta, \; \tilde p_{12} < 0.1\}.
\end{align}
For each combination of the hyperparameters $(m,M,t,s,l)$, we run the Gibbs sampler and estimate the corresponding median and variance for each variable $p_{ij}$. We then choose the combination for which the estimated prior medians satisfy the criteria~\eqref{eq:temp:7304}. If there are multiple solutions, we choose the one that maximizes the total prior variance: $\sum_{i,j} \text{var}(p_{ij})$. 

The heuristic we follow for setting the range of $m$ and $M$ above is as follows: We consider a sequence of i.i.d. random variables $p'_{ij} \sim \text{beta}(\alpha_0,\beta_0)$ for $i=1,\dots,I$ and $j=1,\dots,J$. We then consider the extreme order statistics of this sequence, i.e., the minimum and maximum of $\{p'_{ij}\}$. It is easy to analytically solve for $\alpha_0$ and $\beta_0$ such that the median of the distributions of these two order statistics have specific values. In particular, we solve these equations to match the desired median values $\tilde{p}_{11}$ and $\tilde{p}_{IJ}$.  The values $\alpha_0$ and $\beta_0$ thus obtained provide a good heuristic to set the range $[\min(\alpha_0,\beta_0), \alpha_0 + \beta_0]$ for the grid search on $m$ and $M$. (In fact, we conjecture that the distribution of the extreme order statistics of $\{p_{ij}'\}$ match those of the prior~\eqref{eq:prior} when $\alpha_{ij} = \alpha_0$ and $\beta_{ij} = \beta_0$ for all $i$ and $j$.)

As an example, the set of criteria for in the simulation study \rom{1} is:
\begin{align*}
\{ \tilde{p} \mid \; |\tilde p_{11} - 0.04 | < 0.01 , \; |\tilde p_{44} - 0.34| < 0.01\}.
\end{align*}
We used $n_m = 15$ grid points for each of $m$ and $M$, $n_t = 10$ points for each of $t$ and $s$, and $n_l = 3$ points for $l$.  The grid search resulted in 625 solutions that satisfied the criteria. Out of these, we chose the one that maximizes the total variance (i.e., the trace of the covariance matrix of $\pb$) leading to the following choice of hyperparameters:
\begin{align*}
\begin{array}{clcccl}
\alphab = &  \big[\, 4.52 & 0.4 & \dots & 0.4 &  0.2 \,\big],
\\[1ex]
\betab = & \big[\, 0.74 & 2.23 & \dots & 2.23 &  13.77 \,\big].
\end{array} 
\end{align*}
\section{Acknowledgments}
We would like to thank Arash Ali Amini for helpful comments. This work is supported in part by the National Institute of Health Grant Number R01 CA188480-01A1, the National Center for Research Resources, Grant UL1RR033176, and is now at the National Center for Advancing Translational Sciences, Grant UL1TR000124, 1U01CA232859-01, and P01 CA098912.

\bibliographystyle{rss}
\bibliography{nonparametric_2agents_cleaned}

\end{document}